\documentclass[preprint2]{aastex}
\pdfoutput=1
\usepackage{spr-astr-addons}

\usepackage{bm}
\newcommand{\ppd}{protoplanetary disk}
\newcommand{\mri}{magnetorotational instability}
\newcommand{\scale}[3]{\left(\frac{#1}{#2}\right)^{#3}}
\newcommand{\ut}[1]{\,\mathrm{#1}}
\newcommand{\up}[2]{\,\mathrm{#1}^{#2}}
\newcommand{\percc}{\,\mathrm{cm}^{-3}}
\newcommand{\persc}{\,\mathrm{cm}^{-2}}
\newcommand{\Msun}{\,\mathrm{M}_\odot}
\newcommand{\Lsun}{\,\mathrm{L}_\odot}

\newcommand{\kms}{\,\mathrm{km\,s}^{-1}}
\newcommand{\cross}{\bm{\times}}
\newcommand{\etaH}{\eta_\mathrm{H}}
\newcommand{\etaA}{\eta_\mathrm{A}}

\newcommand{\etaO}{\eta_\mathrm{O}}
\newcommand{\etapara}{\eta_\mathrm{O}}

\newcommand{\curl}{\grad \cross}
\newcommand{\grad}{\nabla}
\newcommand{\divg}{\grad \cdot}
\newcommand{\vv}{\bm{v}}
\newcommand{\J}{\bm{J}}
\newcommand{\B}{\bm{B}}
\newcommand{\E}{\bm{E}}            
\newcommand{\delt}[1]{\frac{\partial #1}{\partial t}}
\newcommand{\delr}[1]{\frac{\partial #1}{\partial r}}
\newcommand{\delz}[1]{\frac{\partial #1}{\partial z}}
\newcommand{\Bh}{\hat{\bm{B}}}
\newcommand{\nH}{n_\mathrm{H}}
\newcommand{\gpersc}{\,\mathrm{g\,cm^{-2}}}
\newcommand{\refitem}{\bibitem{}}
\newcommand{\Mdotscaled}{\dot{M}_{-7}}
\newcommand{\rAU}{r_\mathrm{AU}}
\newcommand{\BonH}{\frac{B_\mathrm{G}}{n_{15}}}

\begin{document}
\title{Magnetic Fields in Protoplanetary Disks}
\shorttitle{Magnetic Fields in Protoplanetary Disks}
\shortauthors{Wardle}
\author{Mark Wardle}
\affil{Department of Physics, Macquarie University, Sydney NSW 2109, 
Australia}
\email{wardle@physics.mq.edu.au}

\begin{abstract}
Magnetic fields likely play a key role in the dynamics and evolution
of \ppd s.  They have the potential to efficiently transport angular
momentum by MHD turbulence or via the magnetocentrifugal acceleration
of outflows from the disk surface.  Magnetically-driven mixing has
implications for disk chemistry and evolution of the grain population,
and the effective viscous response of the disk determines whether
planets migrate inwards or outwards.  However, the weak ionisation of
\ppd s means that magnetic fields may not be able to effectively
couple to the matter.  I examine the magnetic diffusivity in a minimum
solar nebula model and present calculations of the ionisation
equilibrium and magnetic diffusivity as a function of height from the
disk midplane at radii of 1 and 5\,AU. Dust grains tend to suppress
magnetic coupling by soaking up electrons and ions from the gas phase
and reducing the conductivity of the gas by many orders of magnitude.
However, once grains have grown to a few microns in size their effect
starts to wane and magnetic fields can begin to couple to the gas even
at the disk midplane.  Because ions are generally decoupled from the
magnetic field by neutral collisions while electrons are not, the Hall
effect tends to dominate the diffusion of the magnetic field when it
is able to partially couple to the gas, except at the disk surfaces
where the low density of neutrals permits the ions to remain attached
to the field lines.

For a standard population of 0.1\micron\ grains the active surface layers
have a combined column $\Sigma_\mathrm{active}\approx 2\gpersc$ at 1\,AU;
by the time grains have aggregated to 3\,\micron,
$\Sigma_\mathrm{active}\approx 80\gpersc$.  Ionisation in the active layers
is dominated by stellar x-rays.  In the absence of grains, x-rays maintain
magnetic coupling to 10\% of the disk material at 1\,AU (i.e.\
$\Sigma_\mathrm{active}\approx 150\gpersc$).  At 5\,AU
$\Sigma_\mathrm{active} \approx \Sigma_\mathrm{total}$ once grains have
aggregated to 1\,\micron\ in size.

\end{abstract}


\section{Introduction}
\label{sec:intro}
Magnetic fields may efficiently transport angular momentum in \ppd s
either via MHD turbulence driven by the magnetorotational instability
(Balbus \& Hawley 1991; Hawley, Gammie \& Balbus 1994), by
accelerating a wind from the disk surfaces (Blandford \& Payne 1982;
Wardle \& K\"onigl 1993), or by looping above the disk surface and
linking different radii.  Advection and/or stirring by
magnetically-driven turbulence affects disk chemistry (e.g.\ Semenov,
Wiebe \& Henning 2006; Ilgner \& Nelson 2006), the aggregation and
settling of small particles that are the preliminary stages of planet
building (Johansen \& Klahr 2005; Turner et al.\ 2006; Fromang \&
Papaloizou 2006; Ciesla 2007), and magnetic activity at the disk
surface may produce a corona (e.g.\ Fleming \& Stone 2003) and affect
observational signatures of \ppd s.  A magnetically-mediated effective
viscosity would also determine the response of the disk to the
gravitational disturbances of forming planets, modifying the balance
between the torques applied by the outer and inner neighbouring
regions of the disk and therefore the rate and direction of planetary
migration (Matsumura \& Pudritz 2003; Johnson, Goodman \& Menou 2006;
Chambers 2006).  Finally, magnetic fields are likely to play a role in
giant planet formation and the formation and evolution of their
satellite systems (Quillen \& Trilling 1998; Fendt 2003).

One critical uncertainty is the magnitude and nature of the coupling
between magnetic fields and \ppd s, which are very weakly ionised on
account of the self-shielding against ionising sources provided by
their high column density, and by rapid recombinations because
of their high number densities.  In addition the mobility of charged
particles is reduced by the relatively high density, and dust grains
sweep up charges and render them effectively immobile.  Indeed,
magnetic fields were expected to be negligible in protoplanetary
disks, as extremely high diffusion was thought to eliminate the
magnetic gradients responsible for magnetic stresses except at the
disk surfaces (e.g.\ Hayashi 1981).  But these expectations are too
pessimistic.  At any given radius in the disk the density declines
rapidly with height because of tidal squeezing by the gravitational
field of the central star.  The corresponding rapid decline in
diffusivity towards the disk surface is aided by external ionisation
by cosmic rays and by stellar x-rays which dominate cosmic rays by
five orders of magnitude within a few g\,cm$^{-2}$ of the disk surface
(Glassgold, Najita \& Igea 1997; Igea \& Glassgold 1999), so that at
least the surface layers may be magnetically active (Gammie 1996;
Wardle 1997).  In addition, the Hall effect provides a dissipationless
diffusion pathway that can maintain field gradients under a much
broader range of conditions than would otherwise occur (Wardle \& Ng
1999; Wardle 1999; Balbus \& Terquem 2001).  Finally, aggregation and
settling of grains increases the number and mobility of charged
species in the gas phase and reduces diffusivity to the point that
magnetic stresses are important even at the disk midplane (Sano \&
Stone 2002).

Here I explore the nature and magnitude of magnetic diffusion in 
\ppd s through calculations of the ionisation equilibrium in 
a stratified disk exposed to cosmic rays and to x-rays from the central 
star.

\section{Minimum-mass solar nebula}
\label{sec:nebula}
To estimate the degree of coupling of magnetic fields to the matter in
\ppd s we first need a notional model for the physical conditions
within the disk.  I adopt the minimum-mass solar nebula model
(Weidenschilling 1977; Hayashi 1981), in which the surface density of
the solar nebula, $\Sigma$, is estimated by adding sufficient hydrogen
and helium to the solid bodies in the solar system to recover standard
interstellar abundances and spreading this material smoothly in a
disk.  This yields an estimate of the minimum surface density needed
in the solar nebula to form the solar system,
\begin{equation}
    \Sigma\approx 1700 \,\, \rAU^{-3/2}\,\, \gpersc\,,
    \label{eq:Sigma}
\end{equation}
where $\rAU$ is the radial distance $r$ from the Sun in astronomical units.
The disk mass enclosed within $r$ is
\begin{equation}
    M_D = \int_{r_\mathrm{min}}^{r} 2\pi r'\Sigma\,\mathrm{d}r' 
    \approx0.024 \scale{r}{100\ut{AU}}{1/2} \Msun \,,
    \label{eq:Mdisk}
\end{equation}
where $r$ is assumed to be much greater than the inner disk radius
$r_\mathrm{min}$.  The temperature profile of the disk is estimated by
considering the thermal balance for a black body of radius $a$ at a
distance $r$ from the Sun,
\begin{equation}
    \frac{\Lsun}{4\pi r^2}\,\pi a^2 = 4\pi a^2\,\sigma_\mathrm{SB} 
    T^4 \,,
    \label{eq:thermal_equilibrium}
\end{equation}
where $\sigma_\mathrm{SB}$ is the Stefan-Boltzmann constant, which yields
\begin{equation}
    T\approx 280\,\,\rAU^{-1/2}\,\,\ut K.
    \label{eq:T}
\end{equation}
These rough estimates of the distribution of matter and temperature
are broadly consistent with continuum observations of \ppd s around
low mass stars, albeit with resolutions equivalent to $\sim100$\,AU
scales (e.g.\ Kitamura et al.\ 2002; Andrews \& Williams 2005).

The isothermal sound speed in the disk implied by this temperature 
profile is
\begin{equation}
    c_s\approx 0.99\,\,\rAU^{-1/4}\,\,\kms
    \label{eq:cs}
\end{equation}
which is small compared to the local Keplerian speed $v_K$, so the disk is thin
(see eq [\ref{eq:h}] below).  Here I have adopted a mean molecular 
weight of 7/3 $m_p$ appropriate for a mixture of 80\% H$_2$ and 20\% 
He by number.  Toomre's $Q$ parameter (Toomre 1964) is
\begin{equation}
    Q = \frac{c_s \Omega}{\pi G \Sigma}\approx 56
    \,\,\rAU^{-1/4}\,,
    \label{eq:Q}
\end{equation}
where $\Omega = v_K/r$ is the local Keplerian frequency and $G$ is the
gravitational constant, so the disk is not unstable to self gravity
(as expected given $M_D \ll \Msun$).  Of course this represents the
\emph{minimum} solar nebula; the surface density could be increased by
factor of $\sim$30 between 1--100\,AU before the disk becomes
gravitationally unstable (e.g. Durisen et al.\ 2007), although as
noted above large disk masses do not seem to be present in low mass
YSOs.

Assuming that the temperature does not depend on height $z$ above the 
disk midplane, the balance between tidal squeezing by the central
object and its internal pressure imply that the gas
density is stratified as
\begin{equation}
    \rho(r,z) = \rho_0(r)\, \exp\left(-\frac{z^2}{2h^2}\right)\,,
    \label{eq:rho}
\end{equation}
where the disk
scale height $h$ is given by
\begin{equation}
    \frac{h}{r} = \frac{c_s}{v_K}\approx 0.03 \,\,\rAU^{1/4} \,\,.
    \label{eq:h}
\end{equation}
The density at the midplane, $\rho_0$, is determined 
by eq (\ref{eq:Sigma}) and the relation $\Sigma = \sqrt{2\pi}\rho_0 
h$, yielding a number density of hydrogen 
nuclei
\begin{equation}
    \nH = 5.8\times 10^{14} \,\,\rAU^{-11/4}\,\,\percc
    \label{eq:nH}
\end{equation}
at the midplane.
It should be emphasised that the temperature certainly depends on
height because of the effects of external irradiation,
height-dependent dissipation of accretion energy and the effects of
chemistry and optical depth on the ability of a gas parcel to cool
(see e.g.\ Dullemond et al.\ 2007).  Nevertheless the decline in
pressure with height required by tidal confinement of the disk is
provided primarily by the drop in density.

\section{Magnetic field strength}
\label{sec:B}
There are as yet, no direct measurements of the field strength in
protoplanetary disks, apart from remanent magnetism in meteorites
which indicate field strengths in the solar nebula of 0.1--1\,G at
1\,AU (Levy \& Sonett 1978).  Nevertheless the likely range of field
strengths is easily estimated.  A lower limit is established by the
1--10 mG magnetic field strengths measured using Zeeman splitting of
OH 18\,cm lines in the cores of molecular clouds and and in masers in
star forming regions.  This field is likely to be amplified by
compression and shear during the formation of the central star and its
surrounding disk.  It therefore seems likely that
\begin{equation}
    B\ga 10\ut{mG} 
    \label{eq:Bmin}
\end{equation}
in \ppd s.  An upper limit on B is 
provided by equality of magnetic and thermal pressure in the disk midplane,
\begin{equation}
    \frac{B_\mathrm{eq}^2}{8\pi} = \rho_0 \, c_s^2
    \label{eq:equipartition}
\end{equation}
which yields
\begin{equation}
    B \la B_\mathrm{eq}\approx 18\,\,\rAU^{-13/8}\,\,\ut{G} \,.
    \label{eq:Beq}
\end{equation}
Note that the equipartition field $B_\mathrm{eq}$ scales as $\Sigma^{1/2}$ if the disk
surface density is increased over the minimum solar nebula value given
by eq.\ (\ref{eq:Sigma}).  The upper and lower limits are
disparate at 1\,AU but are equal at 100\,AU.

It is worth comparing these limits to the field strength that is
\emph{required} if magnetic fields are responsible for the loss of
angular momentum implied by the observationally inferred disk
accretion rates $ \dot{M}\sim 10^{-7}\Msun\up{y1}{-1}$ (Calvet,
Hartmann \& Strom 2000).  To do this, we take as a starting point the
MHD momentum equation
\begin{equation}
	\rho\, \delt{\vv}+\rho(\vv\cdot\grad)\vv+\grad P + \rho\grad\Phi =
        -\frac{\grad B^2 }{ 8\pi} \,-\, \frac{(\B\cdot\grad)\B}{4\pi}\,,
	\label{eq:Euler}
\end{equation}
where $P$ is the gas pressure and $\Phi$ is the point-mass
gravitational potential of the central star. 
Assuming axisymmetry, and adopting cylindrical coordinates, the
toroidal component of the momentum equation reduces to 
$\rho\, [(\vv\cdot\grad)\vv]_\phi = (\B\cdot\grad\B)_\phi\,/4\pi$. It is 
clear that the options for angular momentum transport under 
these assumptions are limited!  Assuming that $v_z$ is small within
the disk we obtain  
\begin{equation}
	\frac{\rho v_r}{r}\delr{(rv_\phi)} =
	\frac{B_r}{4\pi r}\delr{(rB_\phi)} + 
        \frac{B_z}{4\pi}\delz{B_\phi}\,,
	\label{eq:angmom2}
\end{equation}
which equates the rate of angular momentum loss by the gas to the
magnetic torque.  The magnetic field transports angular momentum
radially and vertically via the $\partial/\partial r$ and
$\partial/\partial z$ terms on the right hand side respectively.  The
LHS is related to the accretion rate at $r$,
\begin{equation}
	\dot{M} = -2\pi r \int_{-s}^{+s} \!\!\!\rho v_r 
        \,\mathrm{d}z\,,
	\label{eq:Mdot}
\end{equation}
by integrating between the lower and upper disk surfaces at $z=\pm s$.
Before integrating eq (\ref{eq:angmom2}) in this manner, note that $v_\phi \approx
v_K \hat{\phi}$ where $v_K = \sqrt{GM/r}$ and use $\divg\B = 0$ to
obtain
\begin{equation}
	\frac{\rho v_r v_K}{2r} \approx \frac{1}{4\pi r^2}\delr{}(r^2 B_r
	B_\phi) + \frac{1}{4\pi} \delz{}(B_zB_\phi) \,.
	\label{eq:angmom3}
\end{equation}
Then integrating vertically between the disc surfaces yields
\begin{equation}
    \frac{\dot{M} v_K}{2r^2}\approx \frac{1}{r^2}
	\delr{}\left(r^2 h <\!\!-B_r B_\phi> \right) - (B_zB_\phi)_s
    \label{eq:Btorque}
\end{equation}
where I have defined
\begin{equation}
	<\!\!-B_r B_\phi> \equiv -\frac{1}{2h}\int_{-s}^s B_r B_\phi\,\mathrm{d}z
	\,.
	\label{eq:BrBphi}
\end{equation}
and $B_zB_\phi = +(B_zB_\phi)_s$ or $-(B_zB_\phi)_s$ at the upper or lower disc 
surfaces respectively.  If $B_r$ and $B_z$ are the same order of magnitude,
and the magnetic flux exiting the disk surfaces is anchored in an
external envelope or outflow, then $B_\phi$ at the disk
surface is non-negligible and the first term in (\ref{eq:BrBphi}) is
smaller than the second by a factor $\sim h/r$.  In this case the
torque per unit area applied to the disk by the net magnetic tension
across it equals the rate of angular momentum loss by the material in
the disk, and
\begin{eqnarray}
\sqrt{|B_z B_{\phi}|_s} \approx
    \left(\frac{\dot{M}v_K}{2r^2}\right)^{1/2}
    \approx 0.2\,\Mdotscaled^{1/2}\,\rAU^{-5/4}\, \ut{G} \,, 
    \label{eq:Brms}
\end{eqnarray}
where $\Mdotscaled$ is $\dot{M}$ in units of $10^{-7}\Msun\up{yr}{-1}$.

This estimate is a robust lower limit on the rms field in the disk, as
it assumes efficient vertical transport of angular momentum by an
ordered magnetic field.  If radial transport dominates instead, then
$<\!\!-B_r B_\phi>^{\!1/2}$ must be larger than the estimate (\ref{eq:Brms}) by a
factor $\sim \sqrt{r/h}$.  If the field is disordered because of turbulence,
then these represent a space \emph{and} time-averaged rms magnetic
field (see, e.g.\ Balbus 2003), and the local field may be somewhat
larger.  For example, simulations of the \mri\ indicate that
$B_\mathrm{rms} \sim 2 <\!\!-B_rB_\phi>\!\!^{1/2}$ (Sano et al.\ 2004).
In any case, one concludes that Gauss-strength fields at 1\,AU are required to
give the inferred accretion rates.

\section{Magnetic diffusion}
\label{sec:diffusion}

Magnetic diffusion is an essential ingredient in any theory of
magnetised astrophysical disks.  For example, field diffusion allows
matter to be accreted while leaving the magnetic field behind.  In
disk-driven wind models magnetic diffusion allows some of the disk
material to be loaded onto field lines and flung outwards from the
disk surfaces (Wardle \& K\"onigl 1993) while the remainder is
accreted.  Diffusion is also associated with energy dissipation and
sets the inner scale of magnetic structures.

The magnitude of the diffusivity is determined by microphysical
processes rather than the wishes of a theorist (or an observer for
that matter!).  In fully-ionised disks the molecular diffusivity is
orders of magnitude too small and an effective turbulent viscosity or
anomalous diffusivity of plasma physics origin has to be invoked as
the origin of the breakdown of ideal MHD. In the weakly-ionised
environment of protoplanetary disks collisions between charged
particles and the dominant neutrals are sufficient to provide the
necessary diffusivity, and may be too efficient, suppressing magnetic
activity, leading to the formation of ``dead'' zones in the disk
(Gammie 1996; Wardle 1997).  If simple neutral-charged particle
collisions are the origin of diffusivity then \emph{quantitative}
calculations can provide a sound physical basis for calculations of
magnetic field evolution in protoplanetary disks.

The finite conductivity of a weakly-ionised gas can be determined by
calculating the drift of charged particles in response to an applied
electric field $\E'$ in the neutral frame, and summing over the
charged species to obtain the current (Cowling 1976).  The drift
velocity $\vv_j$ of each charged species $j$ (mass $m_j$, charge $Z_j
e$) relative to the neutrals (mean mass $m$, density $\rho$) is
determined by balancing the force applied by the electric field, the
magnetic force, and the drag associated with neutral collisions:
\begin{equation}
    Z_j e \E' + Z_j e \, \frac{\vv_j}{c} \cross\B - m_j\gamma_j\rho \vv_j
    = 0 
    \label{eq:vj}
\end{equation}
where $\gamma_j = <\sigma v>_j/(m_j + m)$ and $<\sigma v>_j$ is the
rate coefficient for collisional momentum transfer between species $j$
and the neutrals, so $\gamma_j \rho$ is the collision frequency with
neutrals.  Eq (\ref{eq:vj}) implicitly assumes that the
electromagnetic field and the neutral fluid evolve on a time scale
that is long compared to the inertial time scales of charged
particles, which is generally a good approximation (see Wardle \& Ng
1999).  The direction of the drift of a particle of mass $m_j$, charge
$Z_j e$ is determined by the relative magnitude of the magnetic and
drag terms in eq (\ref{eq:vj}), which is characterised by the Hall
parameter for species $j$
\begin{equation}
    \beta_j = \frac{|Z_j|eB}{m_j c}\frac{1}{\gamma_j\rho} \,
    \label{eq:beta}
\end{equation}
i.e. the ratio of the gyrofrequency and neutral collision
frequency\footnote{Note that Wardle \& Ng (1999) include the sign of
$Z_j$ in the definition of $\beta_j$}.  Note that apart from molecular
factors, the Hall parameter depends only on $B/\nH$.  If $\beta_j \gg
1$ the Lorentz force dominates the neutral drag
\begin{equation}
     Z_j e \E' \approx - Z_j e \, \frac{\vv_j}{c} \cross\B 
    \label{eq:mag_only}
\end{equation}
and the charged particles are tied to the magnetic field lines.  In 
the other limit $\beta_j \ll 1$ the drag dominates
\begin{equation}
    \quad Z_j e \E' \approx \gamma_j m_j \rho\, \vv_j
    \label{eq:drag_only}
\end{equation}
and neutral collisions completely decouple the particles from the
magnetic field.

Inverting eq (\ref{eq:vj}) for $\vv_j$, and forming the current
density $\J=\sum_j n_jeZ_j\vv_j$ yields 
\begin{equation}
    \J = \sigma_O \E'_{\parallel} + \sigma_H \Bh\cross\E'_\perp + \sigma_P 
    \E'_\perp
    \label{eq:J-sigmaE}
\end{equation}
where $\E'_\parallel$ and $\E'_\perp$ are the components of $\E'$
parallel and perpendicular to $\B$ and the ohmic, Hall, and Pedersen
conductivities are (e.g.\ Cowling 1976; Wardle \& Ng 1999):
\begin{equation}
    \sigma_O = \frac{ec}{B}\sum_j n_j|Z_j|\beta_j\,,
    \label{eq:sigmaO}
\end{equation}
\begin{equation}
    \sigma_H = -\frac{ec}{B}\sum_j \frac{n_jZ_j\beta_j^2 }{ 1+\beta_j^2}
    = \frac{ec}{B}\sum_j \frac{n_jZ_j}{ 1+\beta_j^2}\,
    \label{eq:sigmaH}
\end{equation}
and
\begin{equation}
    \sigma_P = \frac{ec}{B}\sum_j \frac{n_j|Z_j|\beta_j }{ 1+\beta_j^2}
    \label{eq:sigmaP}
\end{equation}
respectively, where I have used $\sum n_jZ_j = 0$ in deriving the
second form of $\sigma_H$ in eq (\ref{eq:sigmaH}).

Inverting eq (\ref{eq:sigmaH}) for $\E'$ yields an induction equation of the
form
\begin{eqnarray}
   \delt{\B} =& \hspace{-1em} \curl(\vv\cross\B) - 
   \curl\left[ \etapara\curl\B \right. \nonumber \\
     +  & \hspace{-1em}\,\,\left.\etaH(\curl\B)\cross\Bh    
	 + \etaA (\curl\B)_\perp
    \right]
    \label{eq:induction}
\end{eqnarray}
where 
\begin{equation}
    \eta_O = \frac{c^2}{4\pi\sigma_O}\,,
    \label{eq:etaO}
\end{equation}
\begin{equation}
    \eta_H = \frac{c^2}{4\pi\sigma_\perp}\,\frac{\sigma_H}{\sigma_\perp}\,,
    \label{eq:etaH}
\end{equation}
and
\begin{equation}
    \eta_A = \frac{c^2}{4\pi\sigma_\perp}\,\frac{\sigma_P}{\sigma_\perp}
    - \eta_O
    \label{eq:etaA}
\end{equation}
are the ohmic, Hall and ambipolar diffusivities respectively, where
$\sigma_\perp = \sqrt{\sigma_H^2 + \sigma_P^2}$.

\begin{figure}[t!]
\begin{center}
\vspace*{12pt}
\includegraphics[width=0.35\textwidth]{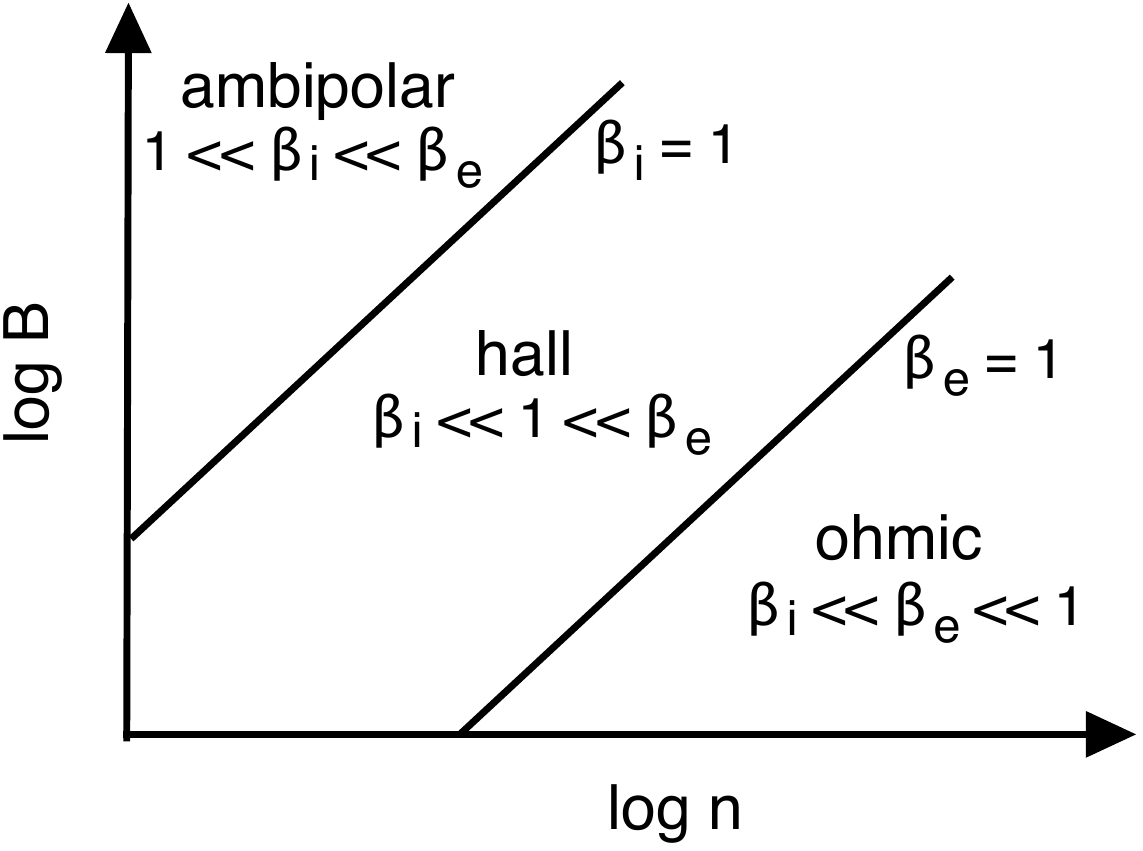}
\vspace*{-12pt}
\end{center}
\caption{The magnetic diffusion regimes of a weakly-ionised,
three-component plasma are determined by the ion and electron Hall
parameters $\beta_i$ and $\beta_e$, which are proportional to $B/n$,
with $\beta_e/\beta_i$ typically $\sim 1000$ (see text).}
\label{fig:diffusion_regimes}       
\end{figure}

If the only charged particles are ions and electrons, then $\etaH =
\beta_e \etaO$ and $\etaA = \beta_i \beta_e \etaO$.  The Hall
parameters for ions and electrons are
\begin{equation}
    \beta_i \approx 4.6\times 
    10^{-3}\,\,\BonH
    \label{eq:beta_i}
\end{equation}
and
\begin{equation}
    \beta_e \approx 3.5 \,\,\BonH \scale{T}{100\ut{K}}{-1/2}
    \label{eq:beta_e}
\end{equation}
respectively, where $B_\mathrm{G}$ is the magnetic field
in Gauss and $n_{15} = \nH/10^{15}\percc$.  As $\beta_e/\beta_i\sim
1000$ there are three distinct diffusion regimes:
\begin{eqnarray}
    \beta_i \ll \beta_e \ll 1 && \textrm{ohmic (resistive)}  
    \nonumber\\
    \beta_i \ll 1 \ll \beta_e  && \textrm{Hall} \\
    1 \ll \beta_i \ll \beta_e && \textrm{ambipolar} \nonumber
    \label{eq:diffusion_regimes}
\end{eqnarray}
which correspond to different regimes of $B/\nH$, as illustrated in
Figure \ref{fig:diffusion_regimes}.  In the ohmic regime, the ion and
electron drifts are unaffected by the magnetic field, in the Hall
regime the electrons are tied to the field but the ions are not, while
in the ambipolar diffusion regime both species are tied to the
magnetic field and drift together through the neutrals.
The presence of grains complicates this picture (Wardle \& Ng 1999),
but generically ohmic or ambipolar diffusion dominates if the majority
of charged particles are tied to the neutrals or the magnetic field
respectively, otherwise Hall diffusion is important.

\begin{figure}[t!]
\begin{center}
\vspace*{12pt}
\includegraphics[width=0.45\textwidth]{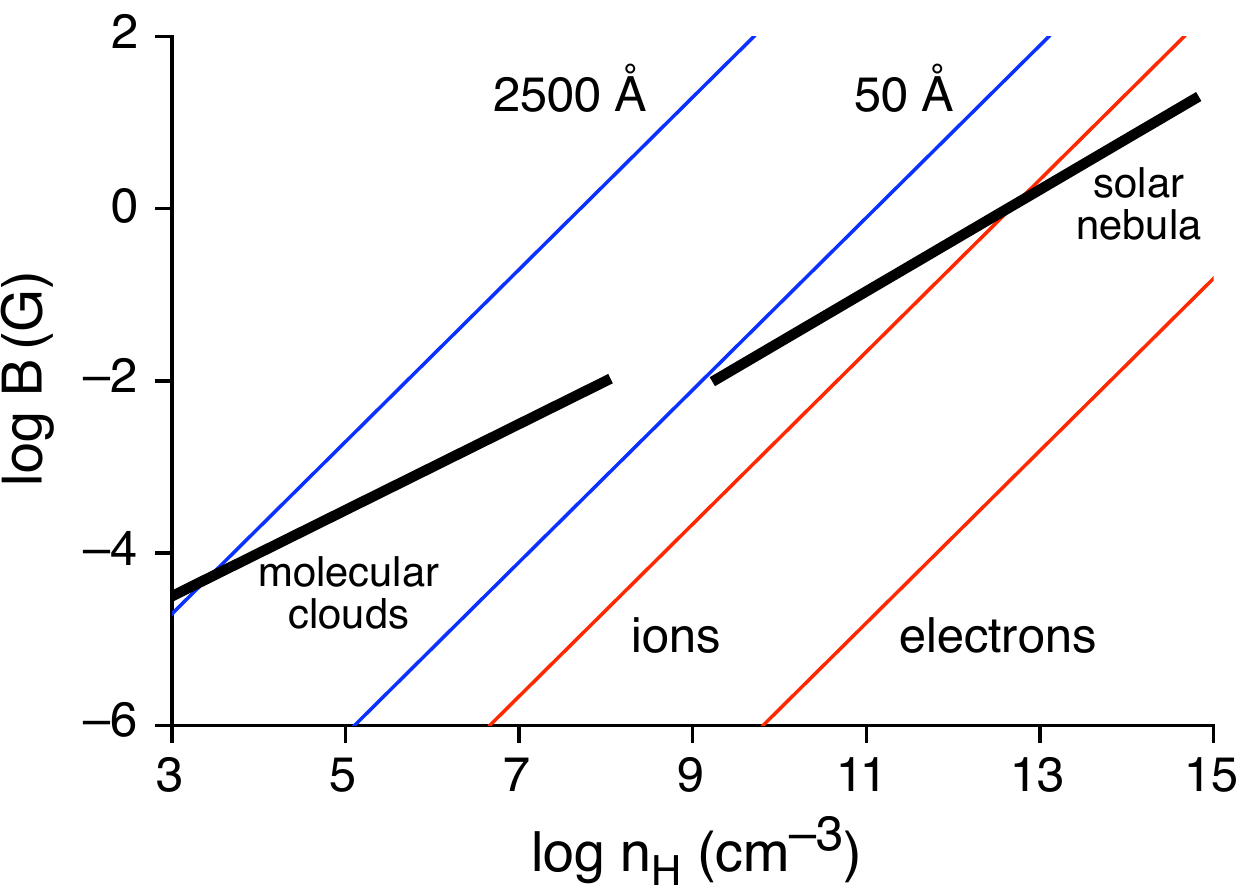}
\vspace*{-12pt}
\end{center}
\caption{Loci in the $\log B$--$\log \nH$ plane where the Hall
parameters are unity for ions and electrons \emph{(red)} and 50 and
2500\,\AA\ grains \emph{(blue)}.  Thick \emph{black} lines indicate
the region occupied by molecular clouds ($B[\mathrm{mG}] \sim
\sqrt{\nH[\percc]}$) and the equipartition field at the midplane of
the minimum-mass solar nebula between 1 and 100\,AU.}
\label{fig:B-n}       
\end{figure}

Which diffusion regimes are relevant?  Figure \ref{fig:B-n} plots the
loci in the $\log B$--$\log \nH$ plane where the Hall parameters are
unity for ions, electrons, and singly-charged grains with radii of 50
and 2500\,\AA. Also shown are the regions occupied by molecular clouds
($B[\mathrm{mG}]\sim \sqrt{\nH[\percc]}$\,) and the midplane of the
solar nebula (eqs [\ref{eq:nH}] and [\ref{eq:Beq}]) between 1 and
100\,AU.  Densities are so
large at the midplane of the solar nebula that ions are decoupled from
the magnetic field by neutral collisions.  This is not so for
electrons except for very weak fields, and so Hall diffusion is
important once grains have settled or aggregated and are not the
dominant charge carriers.  If grains are important they are very much
decoupled from the magnetic field (i.e.\ $\beta_g\ll1$) and ohmic
diffusion dominates.  However, the diffusivity is then so severe that
the magnetic field cannot couple effectively to the gas at all
(Hayashi 1981).

By contrast, in molecular clouds ions and electrons 
are tied to magnetic fields whereas the largest grains are not.  The
charge residing on large grains is usually small except inside shock
waves (Wardle \& Chapman 2006) or at very high densities.  Thus in
molecular clouds, ambipolar diffusion dominates. 

How much diffusion in \ppd s is too much of a good thing?  This depends on the
scale on which one would like the magnetic field to couple to the
matter in the disk.  The weakest criterion for interesting magnetic
effects is to demand that the magnetic field should at least be able
to couple to the Keplerian shear in the disk.  This is required both
by the magnetorotational instability and by disk-driven wind models.
To estimate this, consider the induction equation
(\ref{eq:induction}).  The magnitude of the inductive term 
$|\curl(\vv\cross\B)| \sim
\Omega B$, where $\Omega = v_K/r = c_s/h$ is the Keplerian frequency.
The diffusive term is dominated by vertical gradients, and so
$|\curl(\bm{\eta}\curl\B)| \sim \eta B / h^2$ where $\eta$ is the 
largest of $\eta_O$, $|\eta_H|$ and $\eta_A$.  The diffusion term must be
smaller than the advective term to allow the field to couple to
the Keplerian shear, corresponding to the requirement
\begin{equation}
    \eta \la h c_s
    \label{eq:criterion}
\end{equation}
for good coupling. 
  
\section{Diffusivity calculations}
\label{sec:calculations}

To answer the question of whether the magnetic diffusivity is
too low or high, and which diffusion mechanism dominates in
different regions of \ppd s, we must calculate the ohmic, Hall and
ambipolar diffusivities appearing in eq (\ref{eq:induction}).  These
depend on the abundances of the charged species and their Hall
parameters (see eqs [\ref{eq:etaO}]--[\ref{eq:etaA}]).

To calculate the charged particle abundances at a given radius we
adopt the surface density and temperature of the minimum solar nebula,
and assume that it is isothermal in the $z$-direction.  The ionising
sources are radioactivity (which is largely negligible), cosmic rays,
and the x-ray flux from the central star.  The cosmic ray flux is
assumed to induce an ionisation rate $10^{-17}\up{s}{-1}\up{H}{-1}$
well outside the disk and suffers exponential attenuation with depth
as $\exp(-\Sigma/96 \ut{g}\persc)$ (Umebayashi \& Nakano 1981).  For
the depth dependence of the x-ray ionisation we adopt a fit to the
results of Igea \& Glassgold (1999).  The simple chemical reaction
scheme of Nishi, Nakano \& Umebayashi (1991) and Sano et al.\ (2000)
is adopted, which follows the abundances of the light ions H$^+$,
H$_3^+$, He$^+$, C$^+$, and representative heavier molecular and metal
ions (denoted m$^+$ and M$^+$ respectively), as well as charged
grains.  Their scheme has been extended to include high grain charge
using the cross sections of Draine \& Sutin (1987).  This is necessary
because the temperatures here are high compared to molecular clouds --
the high thermal velocities of electrons can overcome large Coulomb
barriers allowing grains to pick up a high charge through electron
sticking.  Following Umebayashi \& Nakano (1990) I consider
populations of uniform size grains, as well as models in which grains
have settled out.

A sophisticated treatment of the ionisation equilibrium using a
full chemical reaction network has been performed by Semenov, Wiebe \&
Henning (2004), although with a more restricted treatment of grain
charging.

\begin{figure}[ht!]
\begin{center}
\vspace{12pt}
\includegraphics[width=0.45\textwidth]{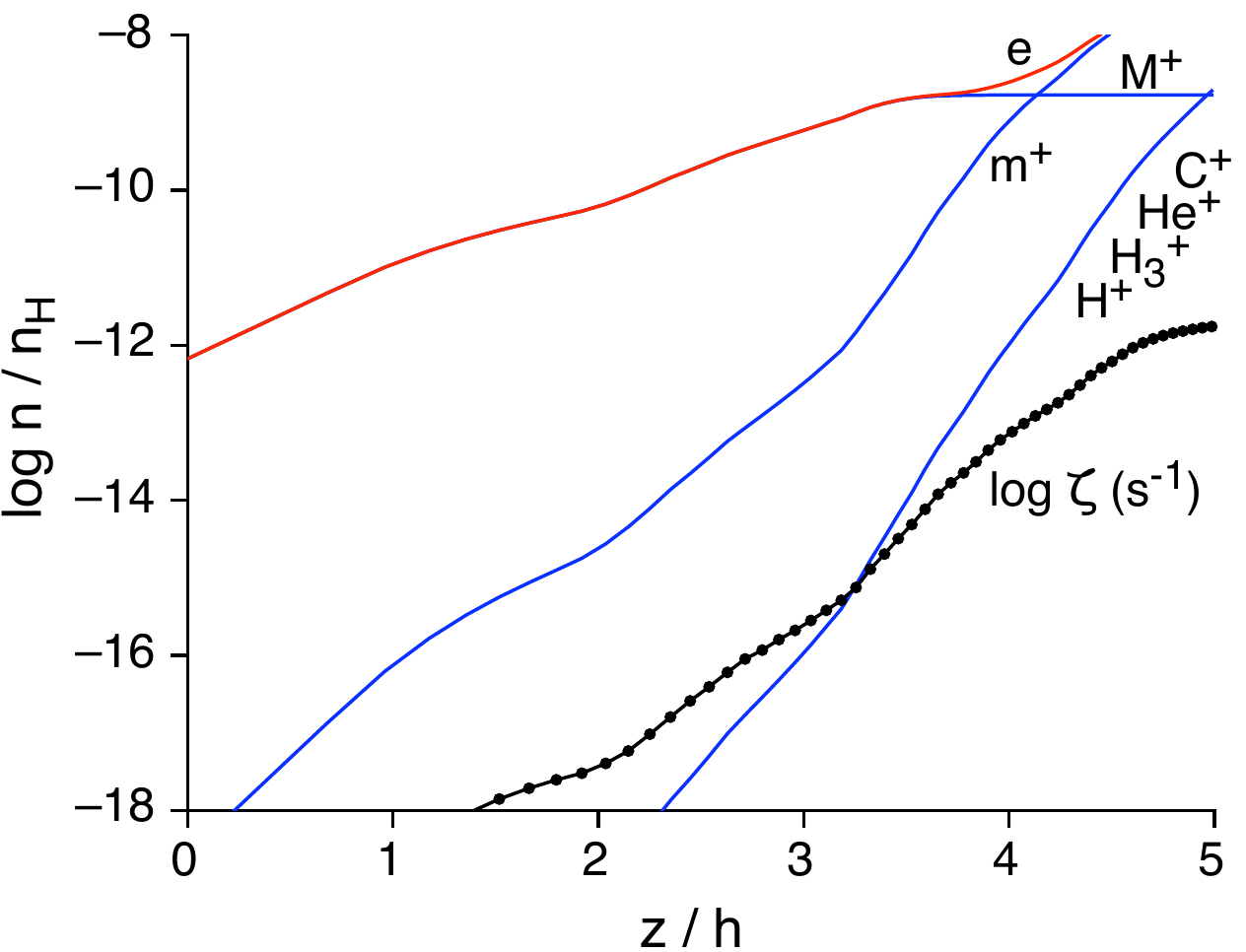}
\vspace{-12pt}
\end{center}
\caption{Ionisation rate \emph{(black dotted curve)} and fractional 
abundance (relative to hydrogen nuclei) of ions 
\emph{(blue)} and electrons \emph{(red)} as a function of height 
above the midplane, $z$, in the minimum solar nebula  
at 1\,AU from the central star. Grains are assumed to have 
aggregated and settled to the midplane.}
\label{fig:x1AU}       
\end{figure}

Figure \ref{fig:x1AU} shows the ionisation rate and abundances at
1\,AU as a function of height $z$ from the midplane in the absence of
grains (i.e. once they have settled).  The ionisation rate is
dominated by x-rays above $z/h = 2$, and by cosmic rays at greater
depths.  The dominant charged species are electrons and metal ions
because the recombination rate coefficient for metal ions is two
orders of magnitude below that for molecular ions.  The ionisation
fraction declines with depth because of the decreasing ionisation rate
and increasing neutral density. 

\begin{figure}[ht!]
\begin{center}
\vspace{12pt}
\includegraphics[width=0.45\textwidth]{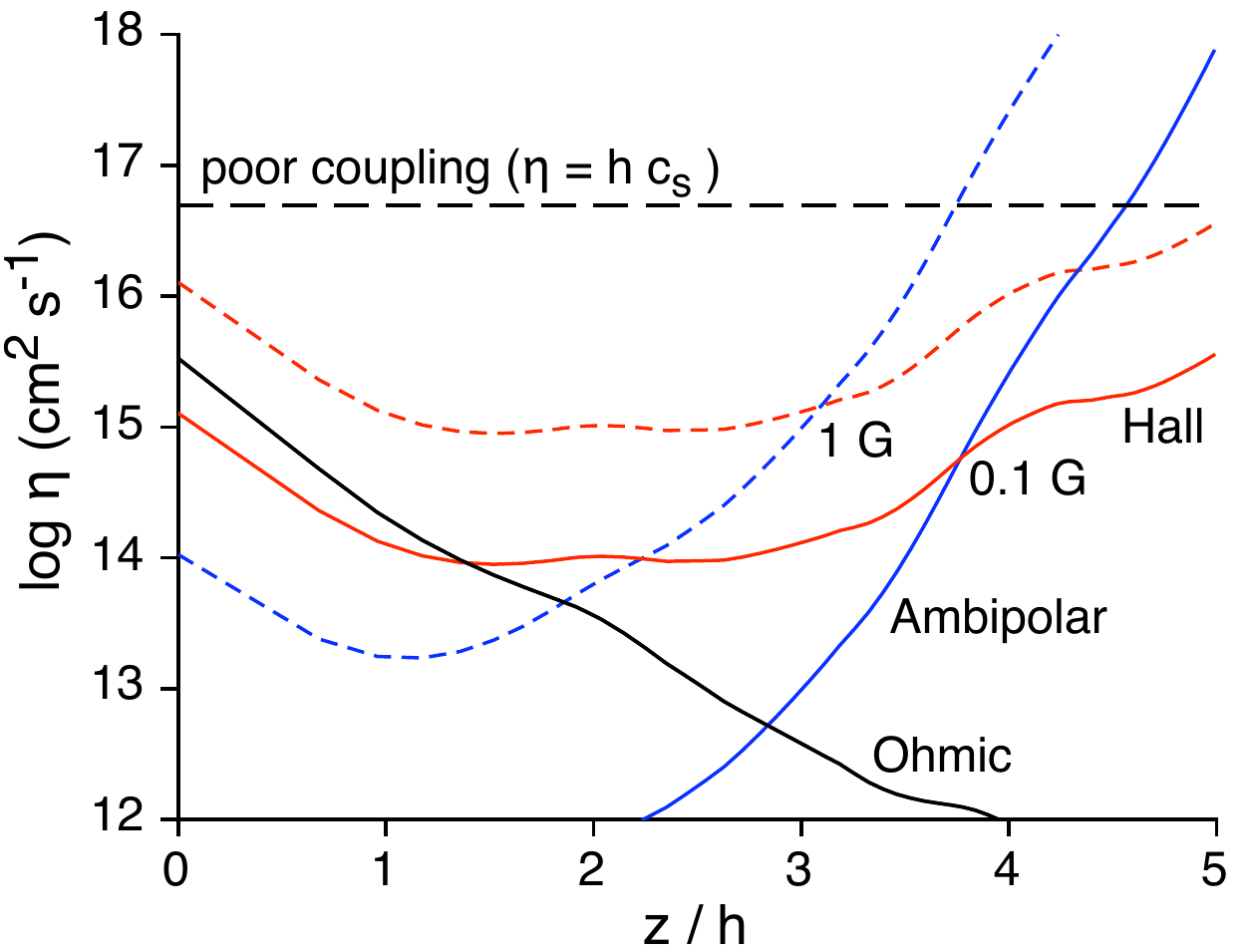}
\vspace{-12pt}
\end{center}
\caption{Magnetic diffusivities (ohmic -- \emph{black} curve; Hall --
\emph{red} curves; ambipolar - \emph{blue} curves) as a function of
height above the midplane derived from the charged particle abundances
plotted in Fig.\ \ref{fig:x1AU} for magnetic field strengths of 0.1
and 1\,G (\emph{solid} and \emph{dashed} curves respectively).  A
single curve for ohmic diffusivity is plotted as this does not depend
on $B$.  The horizontal dashed line indicates the value of diffusivity
at which the magnetic field is unable to effectively couple to the
Keplerian shear in the disk (see \S \ref{sec:diffusion}).}
\label{fig:eta1AU}       
\end{figure}

Figure \ref{fig:eta1AU} shows the corresponding ohmic, Hall and
ambipolar diffusivities as a function of height for field strengths of
0.1 and 1\,G. The ohmic resistivity is independent of the magnetic field
strength, whereas the Hall and ambipolar diffusivities scale linearly
and quadratically with $B$ respectively.  The ohmic diffusivity is
proportional to $n_e$ and declines strongly with height.  The ion and
electron Hall parameters, which are inversely proportional to $\rho$,
increase strongly with height.  As a result, the Hall and ambipolar
diffusivities increase above $z/h\sim 2$ where the neutral density
starts to drop rapidly.  Hall diffusion dominates below 2-3 scale
heights, and ambipolar diffusion dominates at greater heights.  Note
that the magnetic field is able to couple to the shear in the disk below 3--4
scale heights.  Above this, ambipolar diffusion is severe.  At first
sight this is counter-intuitive -- the diffusion is most severe above
the disk where the ionising flux is strongest and the fractional
ionisation is greatest!  The problem is that the \emph{number density}
of charged particles is very low at the disk surface.

\begin{figure}[ht!]
\begin{center}
\vspace*{12pt}
\includegraphics[width=0.45\textwidth]{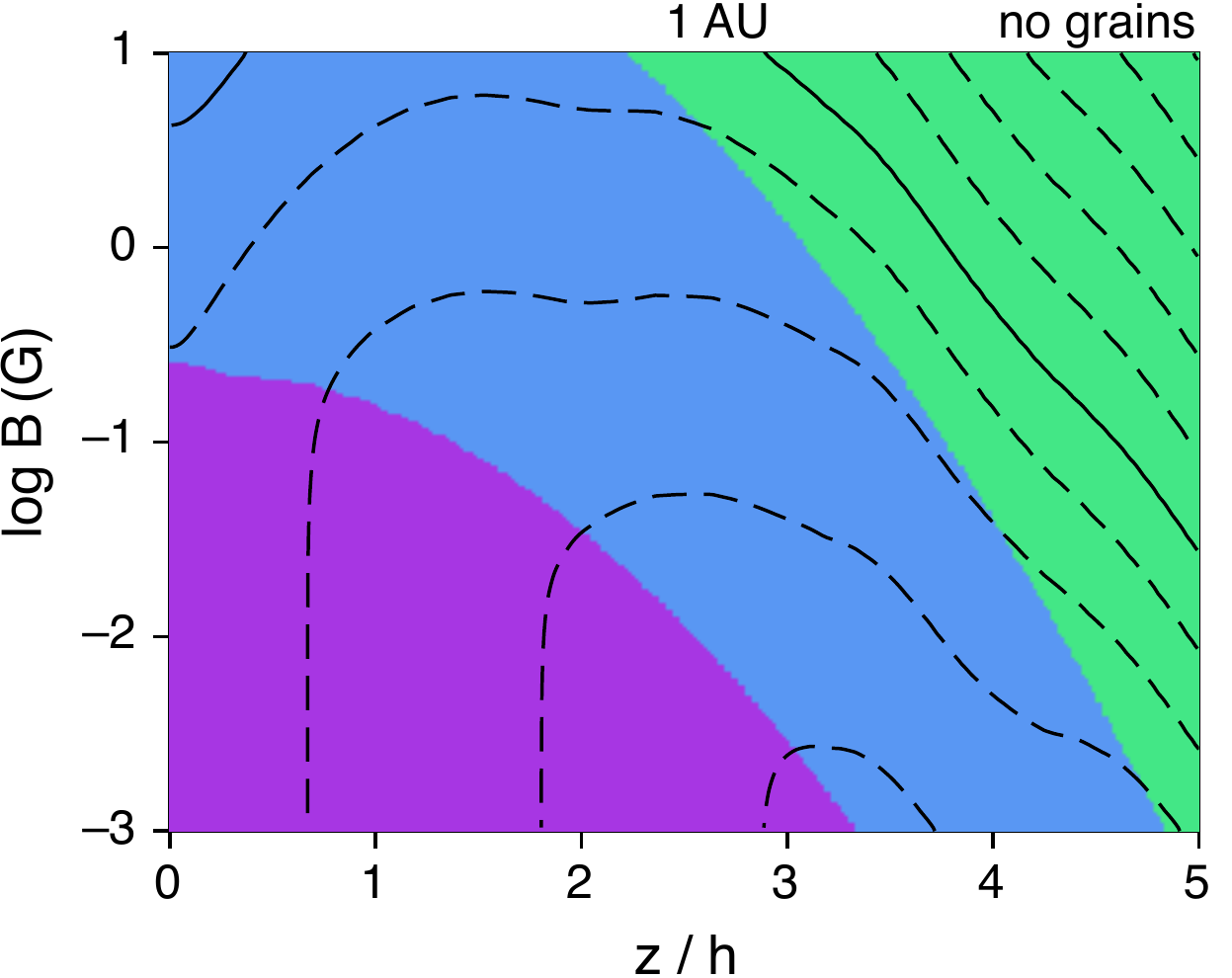}
\vspace*{-12pt}
\end{center}
\caption{Logarithmically-spaced contours of $(\etaO^2
+ \etaH^2 + \etaA^2)^{1/2}$ as a function of height and magnetic field
strength at 1\,AU in the minimum mass solar nebula.  The charged
particle abundances are plotted in Figure \ref{fig:x1AU}.  The contour
levels increase by factors of 10 from $10^{-4}\, h c_s$ at the bottom of
the plot near $z/h=3$ up to $10^5 \,h c_s$ at the very top right-hand
corner of the plot.  The solid contour is the critical value
$hc_s$ -- in the region above this contour the magnetic field cannot
couple effectively to the disk.  The background shading indicates
whether the dominant diffusion mechanism is ohmic \emph{(purple)},
Hall \emph{(blue)} or ambipolar \emph{(green)}.}
\label{fig:Bn1AU}       
\end{figure}

Figure \ref{fig:Bn1AU} extends this plot to other values of $B$ by
representing the diffusivity as contours of $(\etaO^2 + \etaH^2 +
\etaA^2)^{1/2}$ in a $\log B$--$z/h$ plane, with the background colour
indicating the dominant diffusion mechanism.  The curves in Figure
\ref{fig:eta1AU} correspond to horizontal cuts in this plane at $\log
B(G) = 0$ and $-1$.  It is clear that ohmic diffusion is important
only for weak fields near the midplane (i.e. when $\beta_e\la 1$),
whereas ambipolar diffusion is dominant at the disk surface and for
stronger fields (when $\beta_i\gg 1$).  Hall diffusion dominates over
the large range of intermediate conditions.

\begin{figure}[ht!]
\begin{center}
\vspace*{12pt}
\includegraphics[width=0.45\textwidth]{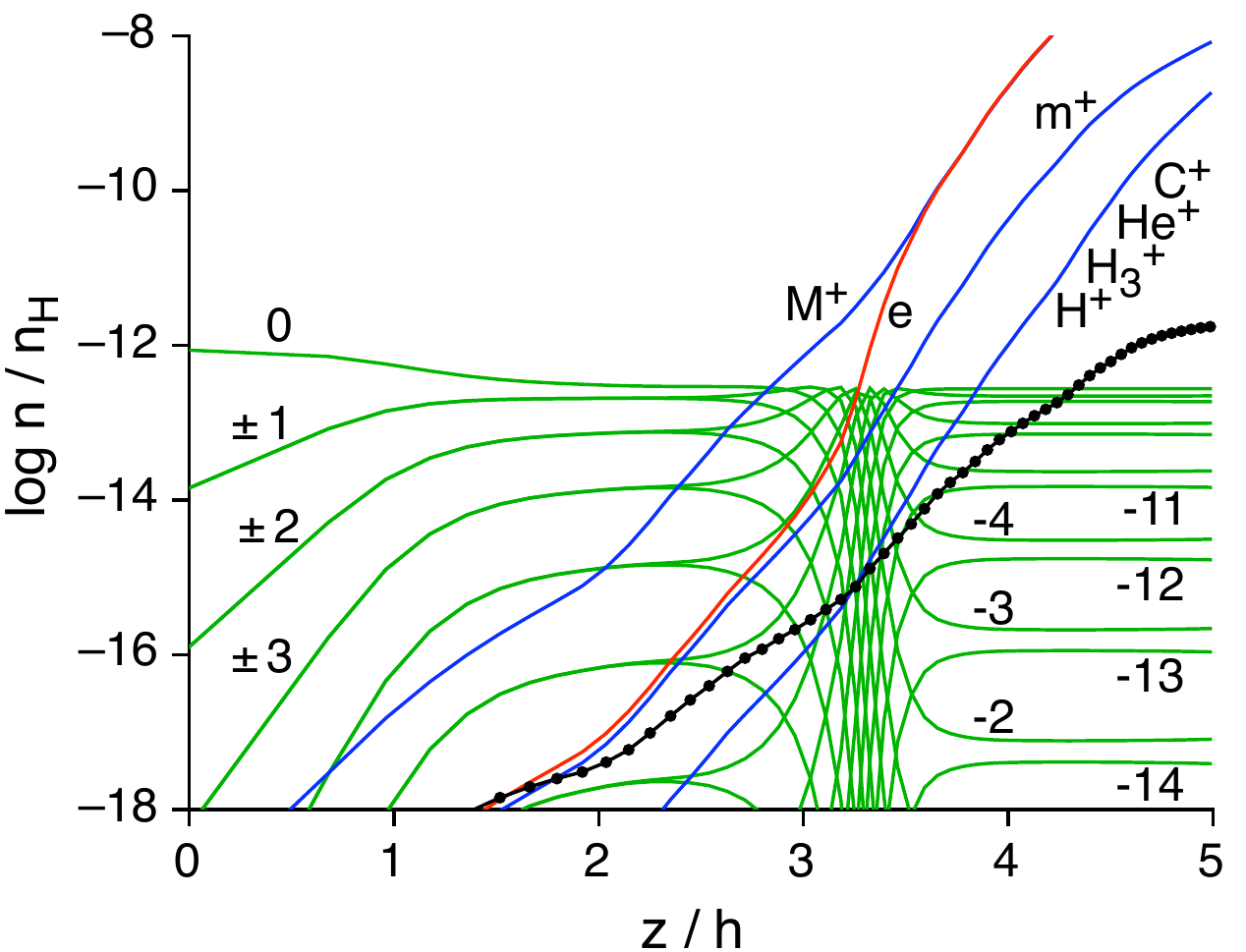}
\vspace*{-12pt}
\end{center}
\caption{As for Fig.\ \ref{fig:x1AU}, but including a standard
interstellar population of 0.1\,$\mu$m radius grains.  \emph{Green}
curves indicate the abundances of different grain charge states $Ze$,
labelled by $Z$.}
\label{fig:xUN1AU}       
\end{figure}

If we add a standard interstellar population of 0.1\micron\ radius
grains, the ionisation equilibrium is very different (Fig.
\ref{fig:xUN1AU}) because the grains acquire a charge via sticking of
electrons and ions from the gas phase.  Above $z/h\approx 4$ the grain
charge is determined by the competitive rates of sticking of ions and
electrons.  The mean grain charge, $\approx-7\,e$, is determined by the
Coulomb repulsion of electrons offsetting their greater thermal
velocity compared to ions.  Most recombinations still occur in the gas
phase.  For $z/h\approx 3$--3.5, the abundances of ions and electrons
have declined to the point that the majority of electrons stick to
grain surfaces before they can recombine in the gas phase, and most
neutralisations occur when ions stick to negatively charged grains.
Closer to the midplane, the ionisation fraction is so low that most
grains are neutral, and ions typically stick to a neutral grain before
encountering a negatively charged grain.  Recombinations mostly occur
in this region through the collision of positive and negative grains.
These regimes were found by Umebayashi \& Nakano (1990), and as expected the
curves are similar to theirs (except for the large grain
charge for $z/h\ga 3$) when the abundances are plotted as a function
of $\zeta/\nH$ rather than $z/h$.

\begin{figure}[ht!]
\begin{center}
\vspace{12pt}
\includegraphics[width=0.45\textwidth]{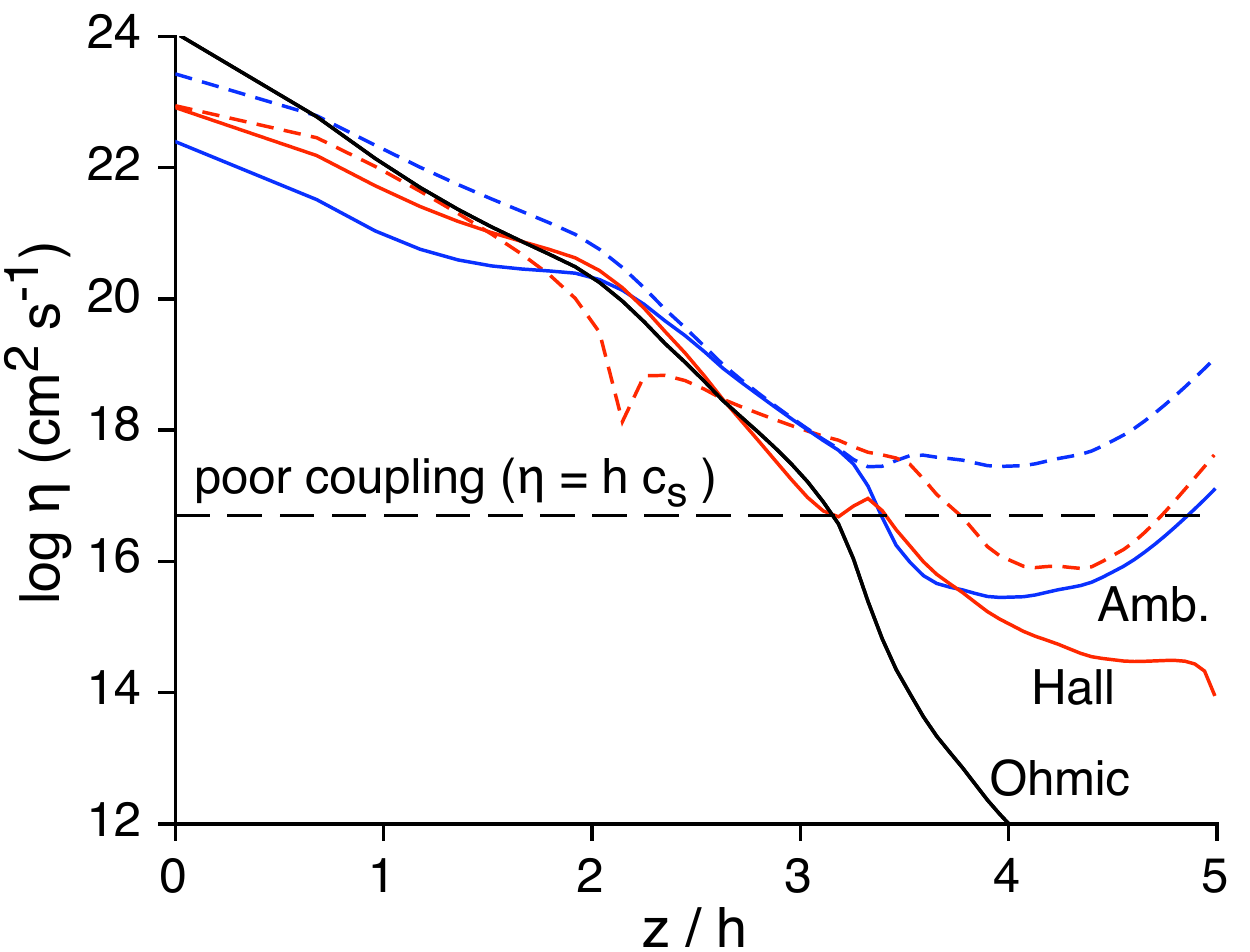}
\vspace{-12pt}
\end{center}
\caption{As for Fig. \ref{fig:eta1AU} but for the charged particle 
abundances plotted in Fig.\ \ref{fig:xUN1AU}.  Note that the vertical 
scale extends to much larger values than that in Fig.\ 
\ref{fig:eta1AU} because the dominance of charged grains for $z/h\la 
3.5$ increases the diffusivity by many orders of magnitude.  The 
kink in the dashed red curve at $z/h\approx 2.2$  occurs because
for $B=1$\,G the Hall diffusivity  becomes negative  
above this point (the absolute value is plotted). }
\label{fig:etaUN1AU}       
\end{figure}

\begin{figure}[ht!]
\begin{center}
\vspace*{12pt}
\includegraphics[width=0.45\textwidth]{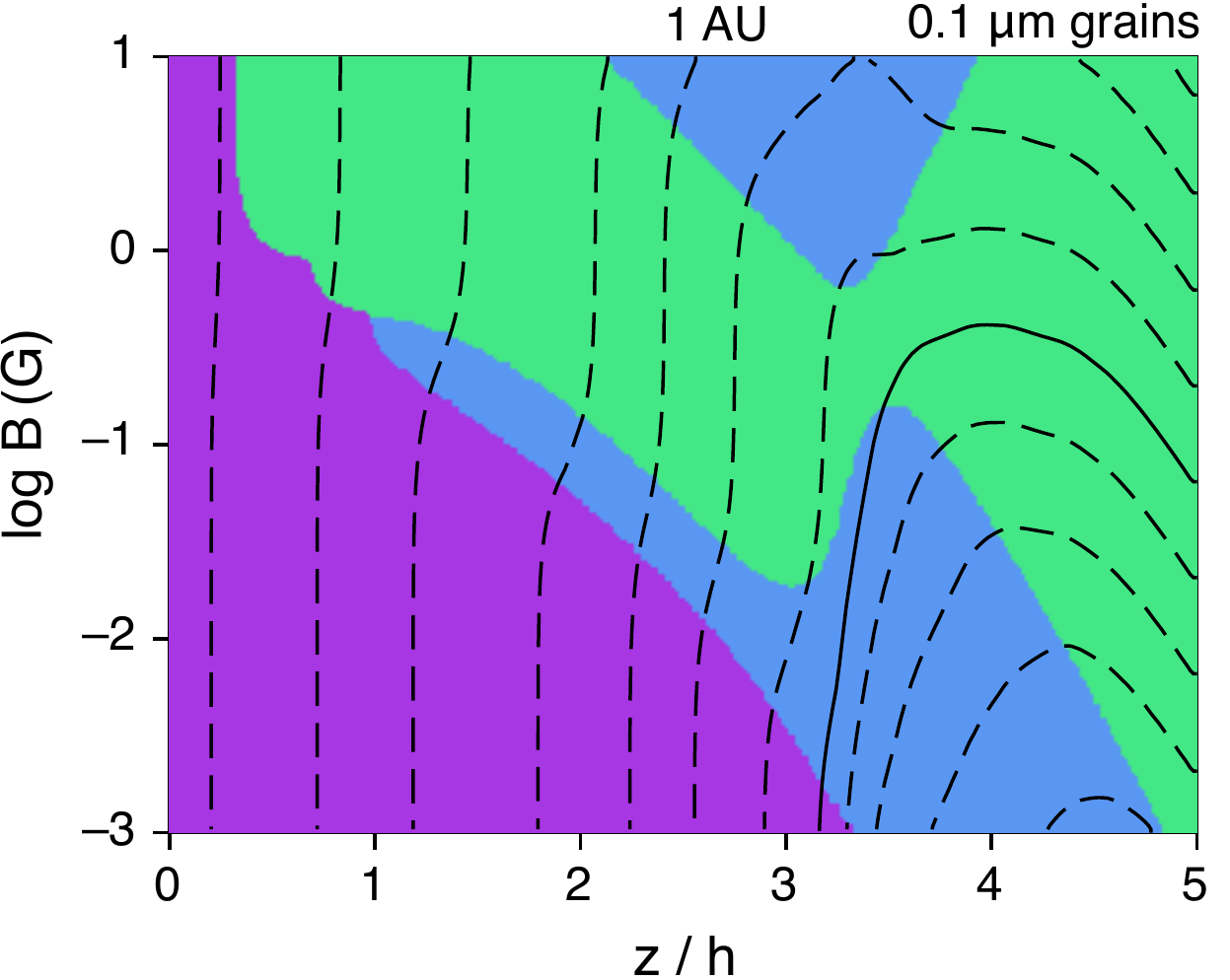}
\vspace*{-12pt}
\end{center}
\caption{As for Fig.\ \ref{fig:Bn1AU}, but for the charged particle 
abundances plotted in Fig.\ \ref{fig:xUN1AU} (i.e.\ including 0.1\,$\mu$m 
radius grains.)}
\label{fig:Bn1AU_01}       
\end{figure}

Grains severely decrease the conductivity by reducing the abundance of
charged particles and because of their much reduced mobility relative
to ions and especially electrons.  Figure \ref{fig:etaUN1AU} shows the
diffusivity as a function of height for field strengths of 0.1 and
1\,G. Close to the midplane the diffusivity is increased by 7-8 orders
of magnitude compared to the case when grains are absent.  It is only
for the weaker field and $z/h\ga 3.5$ that the diffusivity is low
enough to allow the magnetic field to couple to the disk.  The
diffusivity as a function of $z/h$ and field strength is plotted in
Figure \ref{fig:Bn1AU_01}.  The vertical contours on the left hand
half of the plane (i.e.\ for $z/h\la 3$ reflect the strong decline of the
conductivity towards the midplane as the density (and therefore
neutral drag) increases and the number of charge carriers decreases.
For weak to moderate fields the dominant charge carriers -- i.e the
grains -- are decoupled from the magnetic field by neutral collisions
and ohmic diffusion dominates.  The grains become tied to the magnetic
field if it is strong enough and ambipolar diffusion dominates.  Hall
diffusion, which would normally be prominent at intermediate field
strengths, is suppressed by the almost equal numbers of positive and
negative grains.  This is because the Hall effect relies on different
degrees of field line-tying for positive and negatively charged
species.  Above $z/h\approx 3.5$ where grains become unimportant, the
diffusivity behaves as in Figure \ref{fig:Bn1AU}.  The region where
the magnetic field can couple to the differential rotation of the disk
is restricted to $B\la 0.5$\,G and $z/h \approx 3.5$--5.

\begin{figure}[ht!]
\begin{center}
\vspace*{12pt}
\includegraphics[width=0.45\textwidth]{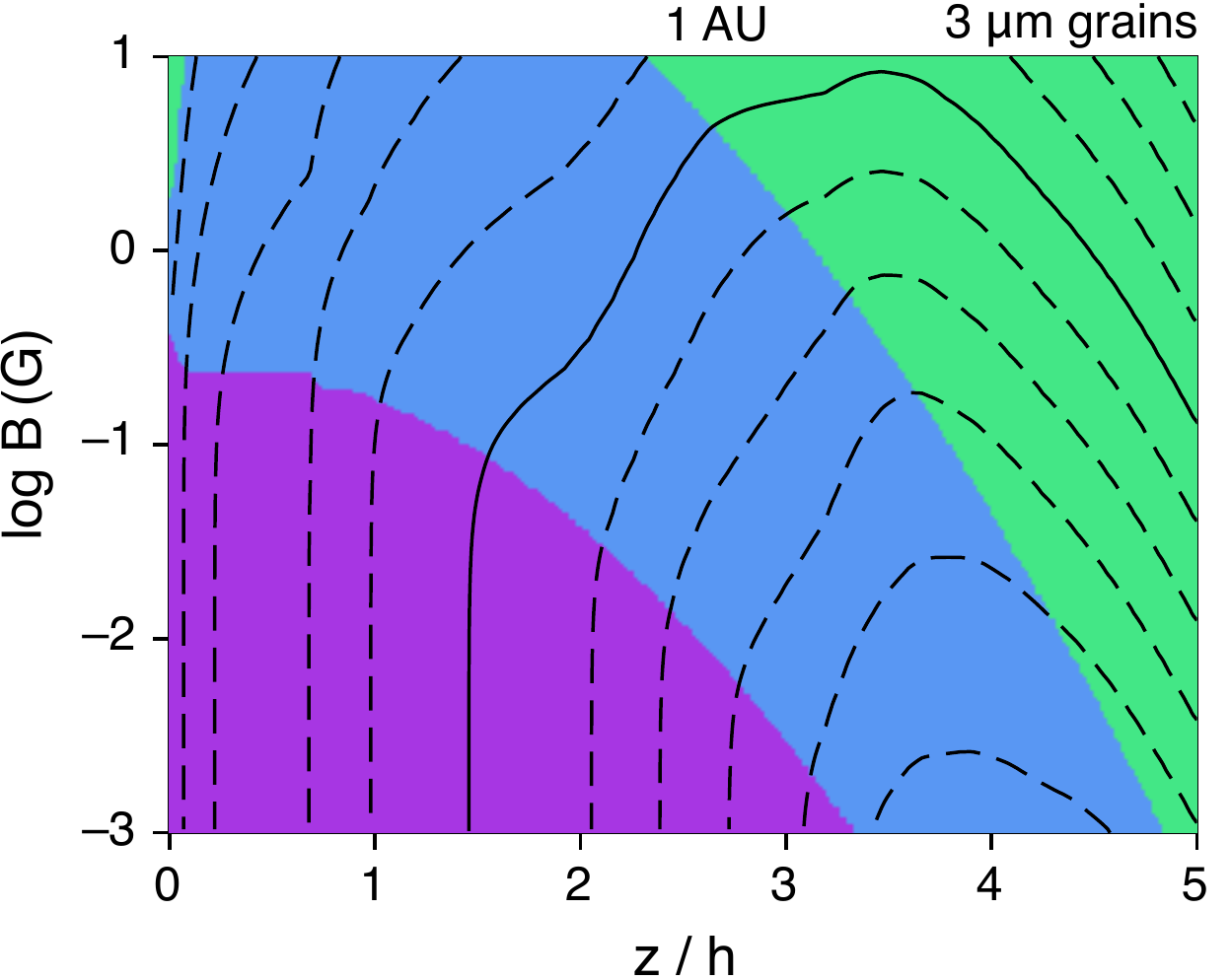}
\vspace*{-12pt}
\end{center}
\caption{As for Fig.\ \ref{fig:Bn1AU}, but assuming that grains have 
radius 3\,$\mu$m.}
\label{fig:Bn1AU_3}       
\end{figure}

These effects can be mitigated by the aggregation of grains, which
reduce the surface area available to capture charged particles from
the gas phase.  For example, I show in Figure \ref{fig:Bn1AU_3} the
diffusivities obtained if grains have aggregated to 3\micron\ in
radius.  This is sufficient to allow the coupled region to extend down
to $z/h\approx1.5$, which is now the height below which the grains
dominate the charged species.  Grains rapidly grow to this size in 
\ppd s, so this suggests that Gauss-strength magnetic fields can couple 
to 10\% of the disk at 1\,AU.

\begin{figure}[ht!]
\begin{center}
\vspace*{12pt}
\includegraphics[width=0.44\textwidth]{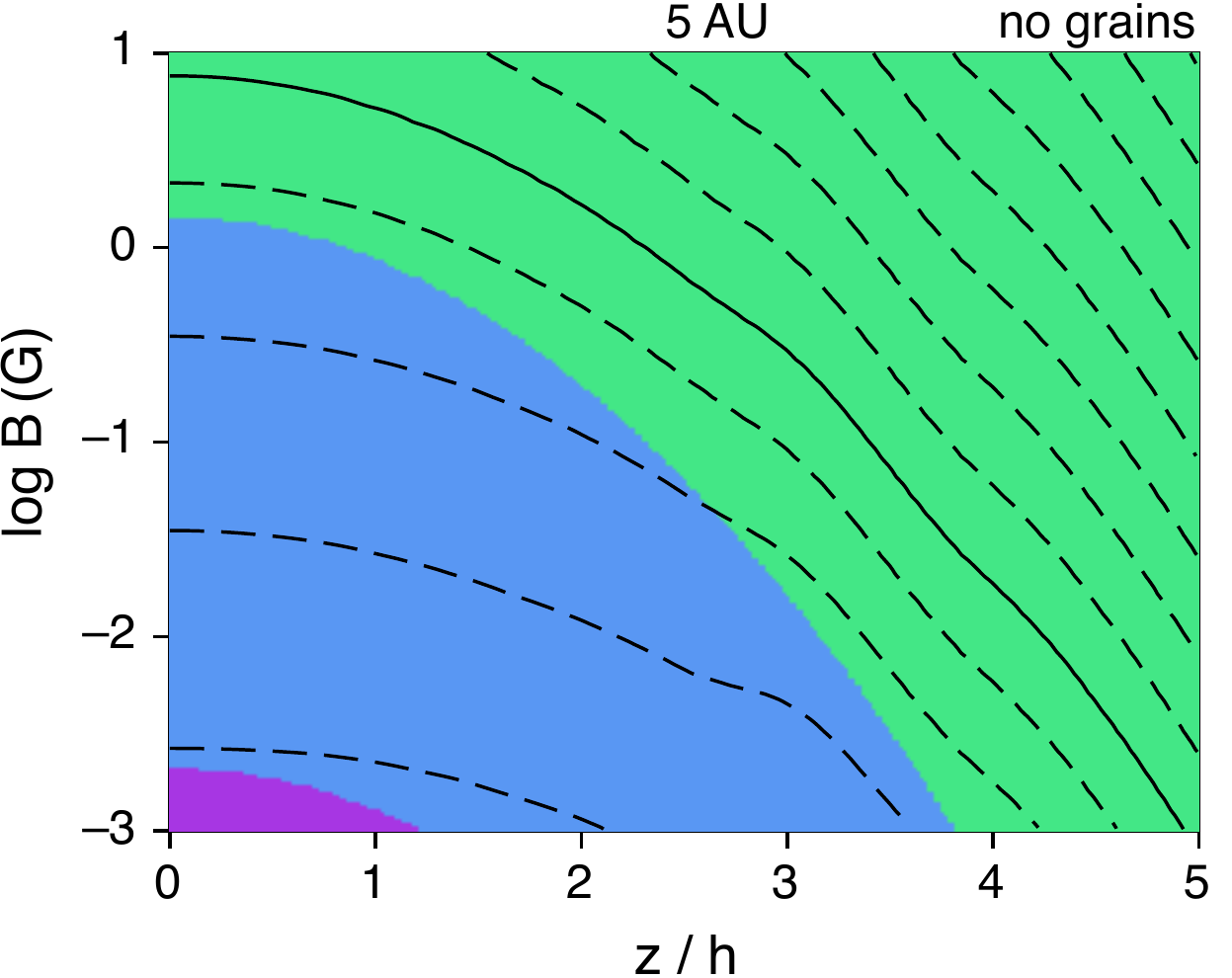}
\vspace*{-12pt}
\end{center}
\caption{As for Fig.\ \ref{fig:Bn1AU}, but at 5\,AU in the minimum 
solar nebula in the absence of grains.}
\label{fig:Bn5AU}       
\end{figure}

At 5\,AU, the region of giant planet formation, the x-ray flux is
reduced by a factor of about 25, while the column density,
$\approx60\ut{g}\persc$, is no longer able to significantly attenuate
the cosmic ray flux impinging on the midplane.  The ionisation rate
drops from $\sim 10^{-13}\up{s}{-1}\up{H}{-1}$ at the surface to $\sim
10^{-17}\up{s}{-1}\up{H}{-1}$ at the midplane.  The ionisation
fraction is similar to that at 1\,AU, but the reduced gas density
allows greater mobility of ions and electrons.  Equipartition fields
($\sim 1$\,G) can easily couple to the midplane in the absence of
grains (see Fig.\ \ref{fig:Bn5AU}).  Note that Hall diffusion tends 
to dominate.

\begin{figure}[ht!]
\begin{center}
\vspace*{12pt}
\includegraphics[width=0.44\textwidth]{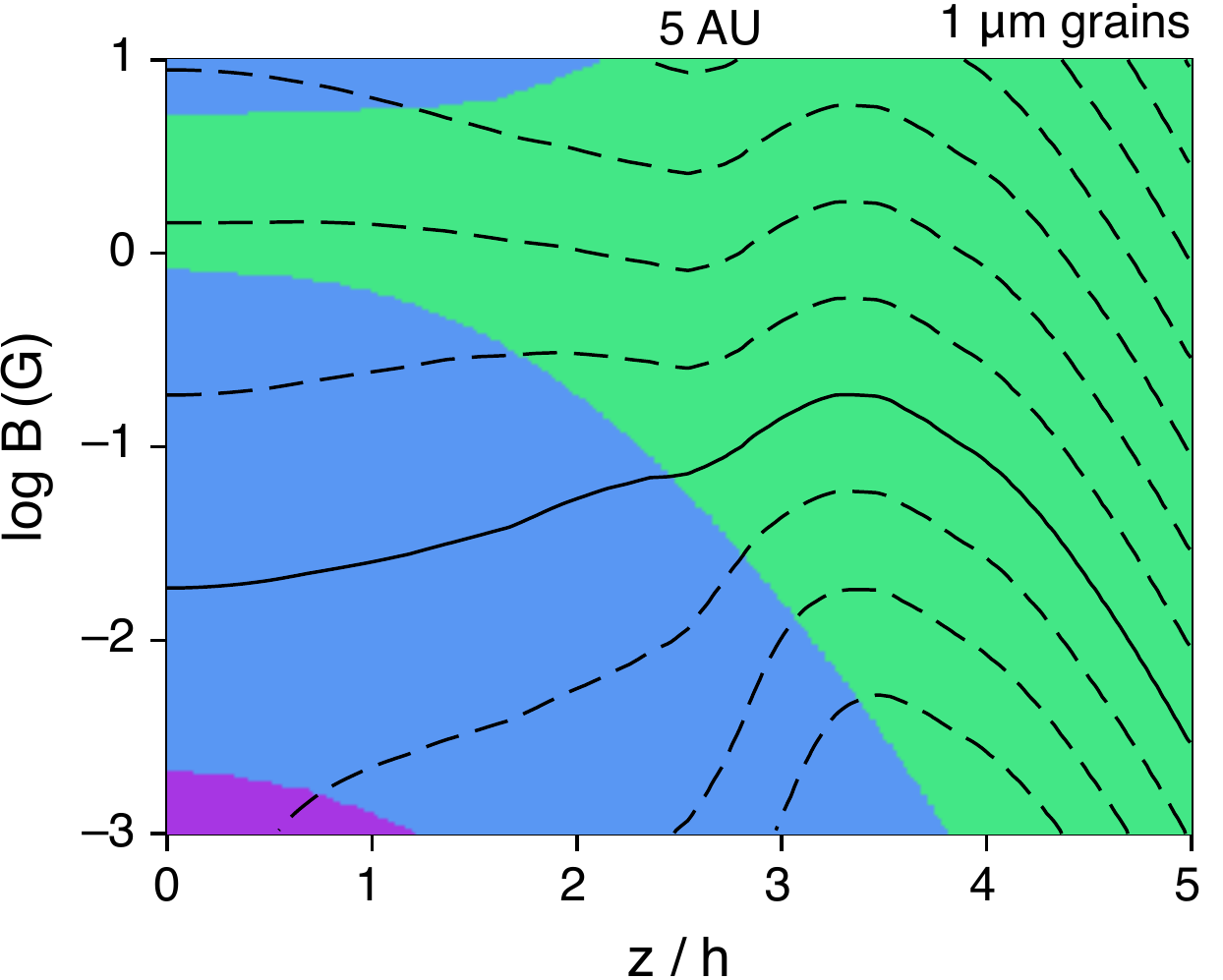}
\vspace*{-12pt}
\end{center}
\caption{As for Fig.\ \ref{fig:Bn1AU}, but at 5\,AU in the minimum 
solar nebula and assuming that grains have 
radius 1\,$\mu$m.}
\label{fig:Bn5AU_1}       
\end{figure}

Grains, if present, increase the diffusion and create a dead zone,
though not as extensive as that at 1\,AU. Increasing the grain size to
just 1\,$\micron$ is sufficient to move the transition from gas-phase
to grain-dominated recombinations down to the midplane.  Then magnetic
fields with strengths below a few tens of milliGauss can couple to the
bulk of the disk (Fig.\ \ref{fig:Bn5AU_1}) and Hall diffusion again is
dominant.

\section{Discussion} 
\label{sec:discussion}

These resistivity calculations assume a minimum-mass solar nebula subject to
cosmic-ray and x-ray ionisation, with grains (if present) assumed to
be of uniform size.  Here I discuss the effect of relaxing these
assumptions.

First, it is not at all clear that \ppd s are subject to ionisation by
interstellar cosmic rays because super-Alfvenic winds may advect them
away from the disk (as noted by Gammie 1996 and Glassgold et al.\
1997), or they may be excluded by unfavourable magnetic geometry such
as hour-glass shaped magnetic field lines vertically
threading the disk or a wound-up field lying in the disk plane.  This
may be compensated by cosmic rays produced by magnetic activity near
the central star or in a disk corona, but it is also unclear whether
these particles are able to impinge on the disk.  If cosmic rays are
absent then the only significant ionisation source are x-rays from the
central star.  Naively one might think that the active layer is then
restricted to the nominal penetration depth of x-rays, i.e. $ \sim
1.4\, m_\mathrm{H}/\sigma_T \sim 3.5\gpersc$.  However, the x-ray
induced active layer is much deeper than this because the unattenuated
x-ray ionisation rate is so high -- x-ray ionisation dominates for
$z/h\ga 2$.  I find that in the absence of dust grains, the active
column density is $\Sigma_\mathrm{active}\approx 150\gpersc$.  When
dust grains are present the active layer lies within the region where
x rays dominate cosmic rays in any case, so this case is unaffected by
eliminating the cosmic rays.

Second, the calculations presented here adopt the surface density for
the minimum-mass solar nebula.  While this is broadly consistent with
observed protoplanetary disks (Kitamura et al.  2002), it is possible
to have surface densities 30-50 times higher without suffering
gravitational instability.  The approximate effect of adding material
to the disk is to multiply up the density everywhere by the same
factor.  The dominant factor in the ionisation equilibrium is the
shielding of the ionising fluxes of cosmic and x-rays by the overlying
column of disk material.  To a first approximation the active
column densities discussed above are unchanged.

Third, I have implicitly used single size grain models to crudely
mimic the effect of aggregation.  Some care is needed in interpreting
these results because of the tendency for the smallest grains to carry
a significant fraction of the grain charge.  For example, although the
typical grain size increases to micron scales on times that are short
compared to the disk evolutionary time scale, the residual small
grains implied by any stochastic aggregation process may still be
important (see, e.g.\ Nomura \& Nakagawa 2006).  The primary role of
grains is to mop up electrons from the gas phase and render them
immobile, increasing the magnetic diffusivity by many orders of
magnitude.  This effect is significant below the transition layer
where the charge density of electrons and grains are comparable (e.g.
at $z/h\approx3.5$ in Fig.\ \ref{fig:xUN1AU}).  Above this transition,
the charge residing on grains is unimportant in determining the
abundances of ions and electrons and their occasional sticking onto
grains yields a mean charge $Z_g(a)\,e \approx -4 kTa/e$ on grains of
radius $a$ (Spitzer 1941; Draine \& Sutin 1987).  The total charge per
unit volume carried by grains with a size distribution $x_g(a)$ (where
the number density of grains with radii between a and $a+da$ is $\nH
x_g(a)\,da$) is therefore $\int \!x_g(a) Z_g(a) \,da$.  The transition
from low to high diffusivity occurs when the charge per hydrogen
nucleus on grains becomes comparable to that in electrons, so the
height of the magnetically active surface layer ($z_\mathrm{mag}$,
say) is a weak, monotonically increasing function of $\int \!a\,x_g(a)
\,da$.  The results presented for the 3\,\micron\ and 0.1\,\micron\
grain models show that $z_\mathrm{mag}/h\sim 2.2$ and 3.5 for $\int
\!a\,x_g(a) \,da = 8.9\times10^{-18}$\,cm and $9.9\times10^{-21}$\,cm
respectively.  By way of comparison, 1\% of the total grain mass
residing in 0.01\micron\ grains would carry approximately the same net
charge as the 0.1\micron\ grain population so a trace residual
population of small grains may poison the magnetically coupled layer.

Finally, although the fractional ionisation increases with height
above the disk the diffusivity also increases.  This is because the
number density of charged particles plummets because the density is
dropping as $\exp(-z^2/2h^2)$.  This potentially affects the ability
to drive a wind from the disk surface.  However it should be noted
that the density profile at large heights becomes increasingly
unreliable due to the effects of temperature stratification and/or the
presence of outflows.

\section{Summary}
\label{sec:summary}

If dust grains have settled to the midplane, or have otherwise been
removed from the gas, the ionisation level in the disk is sufficient
to couple the magnetic field can to the midplane even within 1\,AU of
the central star.  In this case the column density of the magnetically
active region of the disk is $\Sigma_\mathrm{active}\approx
1700\gpersc$.  Hall diffusion dominates (Wardle \& Ng 1999; Sano \&
Stone 2002) because the ions are decoupled form the magnetic field and
electrons are not.  Notably, the Hall effect imparts an implicit
handedness to the fluid dynamics, which becomes sensitive to a global
reversal of the magnetic field direction.

Dust grains, if present, substantially increase magnetic diffusion by
soaking up electrons and ions from the gas phase.  Grains have huge
cross sections for neutral collisions and this combined with their
large inertia effectively decouples them from the magnetic field and
the fluid becomes resistive.  Magnetic diffusion is so severe that the
field is unable to couple to the shear in the disk except close to the
disk surface.  For a standard interstellar population of
0.1\micron-radius grains the active surface layers have a combined
total column $\Sigma_\mathrm{active}\approx 2\gpersc$.  As grains
aggregate, their effect on the ionisation equilibrium is reduced, and
by the time grains have aggregated to 3\,\micron,
$\Sigma_\mathrm{active}\approx 80\gpersc$.  At 5\,AU the lower column
density and reduced gas density and shear means that the coupling is
more easily maintained.  In this case $\Sigma_\mathrm{active} \approx
\Sigma_\mathrm{total}$ once grains have aggregated to 1\,\micron\ in
size.

If a substantial grain population is present, ionisation in the
magnetically active layers is dominated by x-rays so that there is
little change if cosmic rays are unable to impinge on the disk.  In
the absence of grains and cosmic rays, stellar x-rays maintain
magnetic coupling to 10\% of the disk material at 1\,AU (i.e.\
$\Sigma_\mathrm{active}\approx 150\gpersc$) and almost all of it at
5\,AU once grains have aggregated or been removed by settling, which 
is expected to happen on a time scale of a few thousand years (e.g.\ Nomura \&
Nakagawa 2006).

\newcommand{\ARAA}{Ann.\ Rev.\ Astron.\ Astrophys.}
\newcommand{\AnA}{Astron.\ Astrophys.}
\newcommand{\ApJ}{Astrophys.\ J.}
\newcommand{\ASS}{Astrophys.\ Space Sci.}
\newcommand{\MNRAS}{Mon.\ Not.\ Roy.\ Astron.\ Soc.}
\newcommand{\PASJ}{Publ.\ Astron.\ Soc.\ Jpn.}

\end{document}